\documentstyle[preprint,aps]{revtex}
\begin{document}
\draft
\title{{The Bulk Josephson response of the d-g-wave
cuprate superconductor}}
\author{\Large M.Yu.Kuchiev, P.V.Shevchenko$^{a}$, O.P.Sushkov$^{b}$}
\address
{School of Physics, The University of New South Wales,
Sydney 2052, Australia}
\date{20 May 1997}
\maketitle
\begin{abstract}
We consider the implications of the small Fermi surface
for the ab-plane microwave absorption. The small Fermi
surface results in two superconducting condensates.
Linear combinations of these condensates correspond to the d- and
g-wave pairings. 
Microwave electric field applied in the
ab-plane induces Josephson transitions between the
condensates. Dependence of the  absorption
upon direction, amplitude and frequency of the
external microwave field is calculated.
Results of the calculation are compared with the
observed recently novel non-linear phenomena in 
microwave absorption of $YBaCu_2O_{7-\delta}$ 
single crystals.
\end{abstract}
\bigskip
Corresponding author: Pavel Shevchenko,\\
School of Physics, 
The University of New South Wales, Sydney, 2052, Australia\\
ph.(612)9385 61 32 fax(612)9385 60 60 
e-mail: pavel@newt.phys.unsw.edu.au
\bigskip\bigskip

\pacs{PACS codes: 74.50.+r, 74.25.Nf\\
keywords: nonlinear absorption in two-gap superconductors}

\section{Introduction}
Recently a coherent Josephson response of the entire
ab-plane has been observed in the microwave impedance of 
$YBaCu_2O_{7-\delta}$ single crystals\cite{Zhai}. 
This observation can be considered as a direct indication of the 
two-condensate nature of the cuprate superconductivity.
The purpose of the present paper is to describe the coherent
microwave Josephson response within the theory of high-temperature
superconductivity based on the t-J model and the assumption of a small
Fermi surface. 

There is a controversy in literature about the shape of the Fermi surface 
in cuprate superconductors.  In the early days it was believed that 
it is the small Fermi surface of the doped Mott insulator \cite{Bed}. 
Later many of the results from photoelectron spectroscopy (PES) have
been interpreted as favoring the large Fermi surface in agreement with
the Lattinger's theorem \cite{Dessau}. On the other hand the most recent PES
data\cite{Aebi,Chak,Wells,Mars} once more give indications of the small Fermi
surface for underdoped samples.
In the present paper we consider the doped Mott insulator
scenario with the small Fermi surface consisting of hole pockets 
around $(\pm \pi/2, \pm \pi/2)$, see Fig.1a.
It is widely believed that the $t-J$ model
describes the main details of the doped Mott insulator. To fit the 
experimental hole dispersion one needs to extend the model 
introducing additional hopping matrix elements 
$t^{\prime}$, $t^{\prime \prime}$ 
(see. e.g. Refs.\cite{Naz,Bala,Sus}), but basically it is the 
$t-J$ model.  Superconducting pairing  induced
by the spin-wave exchange in the $t-J$ model
has been considered in the papers \cite{Flam,Bel}.
It was demonstrated \cite{Flam}  that there is an infinite set of
solutions for the superconducting gap and  all the solutions have nodes
along the lines $(1,\pm 1)$, see Fig.1a. (It is very convenient to
use the magnetic Brillouin zone, but it can certainly  be mapped
to the full zone.) Using translation by the vector of the inverse magnetic lattice
the picture can be reduced to two hole pockets centered around
the points $(\pm \pi/2, \pi/2)$, see Fig.1a. The superconducting pairing is
the strongest between particles from the same pocket, and the lowest energy
solution for the superconducting gap has only one node line in each pocket.
Having this solution in a single pocket, one can generate two solutions in
the whole Brillouin zone taking symmetric or antisymmetric combinations
between the pockets. The symmetric combination corresponds to the d-wave
(Fig.1b), and the antisymmetric combination corresponds to the g-wave (Fig.1c)
pairing. A possibility of generating new 
solutions by taking different combinations between pockets was first 
demonstrated by Scalettar, Singh, and Zhang in the paper \cite{Scal}.
The g-wave pairing in cuprate superconductors has also been  
considered by Buchholtz, Palumbo, Rainer, and Sauls\cite{Buch}.
They discussed the most general case of two d-wave order parameters
and two g-wave order parameters and their behavior near the surface of
the superconductor. 
The small Fermi surface case considered in
the present paper is different, because the pairing exists only in one d-wave 
channel and in one g-wave channel. On the other hand, these two pairings
are almost degenerate and therefore the new spatial scale $l_{\gamma}$
arises. This scale is much larger than the superconducting correlation 
length $\xi$.
The energy splitting between the d- and g-wave solutions in the $t-t^{\prime}-J$
model has been investigated numerically in the Ref.\cite{Bel}.
The g-wave solution disappears, and only the d-wave one survives,
as soon as the hole dispersion is degenerate along the face of the magnetic
Brillouin zone. In this situation there are no pockets
and one has a large open Fermi surface for arbitrary small hole 
concentration.
However for small well separated pockets the d- and g-wave solutions are almost
degenerate. In the pure $t-J$ or $t-t^{\prime}-t^{\prime \prime}-J$ model the
d-wave solution always has the lower energy. However if one extends the model
including the nearest sites hole-hole Coulomb repulsion, the order can be
inverted. The nearest sites repulsion does not influence the  g-wave pairing 
and substantially suppresses the d-wave pairing. So it is possible that 
the real ground state has the g-wave superconducting gap.
We would like to note that the g-wave ground state does not contradict the
existing experimental data on the Josephson tunneling \cite{Harl}. In this 
case the tunneling current pumps the
g-wave into d-wave in a thin layer near the contact, giving the
interference picture which is very close to that for a pure d-wave ground 
state\cite{stat1}.
For the present work it does not matter  whether the
ground state has d-wave or g-wave symmetry. What is important is that there are
two coupled superconducting condensates.
Then the idea is straightforward: external microwave field induces Josephson
transitions between these condensates. We stress that the condensates
are not separated in coordinate space, they coexist at each point and are 
separated only in k-space.

 The possibility of two close condensates has been suggested
for conventional superconductors a long time ago by 
Suhl, Matthias and Walker \cite{Walker}. Collective excitations 
in such a system were considered by Legget \cite{Legget}, and then
Rogovin discussed the behavior of the two component system in the
external microwave electric field \cite{Rog}.
In the present work we consider the response of the cuprate superconductor
to an external microwave electric field parallel to the ab-plane.
Small orthorhombic distortion is not important for this process, and therefore 
it is  neglected. Hence we neglect twinning and consider a single crystal.

\section{The Ginzburg-Landau Lagrangian and free energy}

According to the microscopic picture described above we should consider
simultaneously the d- and g-wave pairings. Let us formulate an effective
Ginzburg-Landau theory describing this situation.
In the first approximation we can neglect the interaction between pockets
in k-space. Then  half of the holes belong to one pocket and the rest
belong to the other pocket,
and we should introduce two macroscopic condensates corresponding to the
pockets,
$\Psi_1=\left|{\Psi_1}\right|\mbox {e}^{i\phi_1}$,
$\Psi_2=\left|{\Psi_2}\right|\mbox{e}^{i\phi_2}$,
where $\left|{\Psi_1}\right|=\left|{\Psi_2}\right|=\sqrt{N_h}/2$,
$N_h$ is the number density of condensate holes.
The effective Lagrangian of the system in an external electric field $E$
is of the form (hereafter we set $\hbar=1$)
\begin{equation}
\label{L}
L=\sum_{p=1,2}\frac{i}{2}\left(\Psi_p^* \dot\Psi_p-\Psi_p \dot\Psi_p^* \right)
-F
\end{equation}
where $F$ is Ginzburg-Landau free energy
\begin{equation}
\label{F}
F=\int\left\{\sum_{p=1,2}\left({1\over{2m_p^\alpha}}|\nabla_\alpha\Psi_p|^2
-a\left|\Psi_p\right|^2 +b\left|\Psi_p\right|^4
+2e\varphi\left[\left|\Psi_p\right|^2- N_0/4\right]\right)
+{{E^2}\over{8\pi}}\right\}dV+F_{\mbox{{\small int}}},
\end{equation}
where $\varphi$ is a scalar potential, $N_0$ is an equilibrium number density 
of condensate holes, and $F_{\mbox{{\small int}}}$ is a small interaction 
between pockets. We choose axes x and y to be directed at $45^o$ with respect
to the crystal axes a and b. For such choice the mass tensor in each pocket is 
diagonal and $m_p^{\alpha}$, $\alpha=x,y$ are corresponding masses.
Due to symmetry of the pockets $m_1^x=m_2^y=m$ and $m_1^y=m_2^x=M$.
In eq. (\ref{F}) and hereafter we assume summation over $\alpha=x,y$.
Interaction between the pockets is described by term $F_{\mbox{\small int}}$ 
in eq.(\ref{F}). Following Legget \cite{Legget} we use the simplest form of 
this 
interaction 
\begin{equation}
\label{Fint}
F_{\mbox{{\small int}}}=
\gamma \int \left(\Psi_1^*\Psi_2+\Psi_1\Psi_2^*\right)dV ,
\end{equation}
where $\gamma\ll a$ is a small parameter of the interaction. 
For the homogeneous case\\
 $F_{\mbox{{\small int}}}$ $\to$
$2\gamma V \left|\Psi_1\right|\left|\Psi_2\right|\cos(\phi_1-\phi_2)$,
where $V$ is the total volume. The Hamiltonian corresponding to the 
Lagrangian (\ref{L}) is the Ginzburg-Landau free energy (\ref{F}).
The equilibrium values of the order parameters are
\begin{equation}
\label{psi}
\left|\Psi_1\right|^2=\left|\Psi_2\right|^2=\left|\Psi\right|^2=
\frac{a+\left|\gamma\right|}{2b}={N_0\over 4}\approx \frac{a}{2b} \ \ .
\end{equation}
The ground state phase difference $\phi= \phi_2 -\phi_1$
is determined by the sign of $\gamma$. If $\gamma>0$ then 
$\phi=\pi$ and the ground state pairing has
g-wave symmetry. If $\gamma<0$ then $\phi=0$ and  the ground 
state pairing has d-wave symmetry.

Note that the free energy (\ref{F}) is written in the long-wavelength
limit. This approximation is valid if the wavelength of the microwave 
field  is much larger than the superconducting correlation length 
$\xi=\left(2a\sqrt{m M}\right)^{-1/2}$ as well as the scale related to the 
weak coupling $l_{\gamma}=\left(4\gamma\sqrt{m M}\right)^{-1/2}$: 
$\lambda \gg l_{\gamma},\xi$. In this limit it is convenient to use a gauge 
with the scalar potential of the external field $\varphi=-{\bf E}\cdot{\bf r}$
and the vector potential ${\bf A}={1\over 2}{\bf H} \times {\bf r}$.
In this gauge the effect of the vector potential is small and therefore it is
neglected in the free energy (\ref{F}). Note also that there is an obvious
upper limit on the applied electric field: $eE\xi \ll \Delta$, where
$\Delta \sim 300K$ is a typical value of the superconducting gap.

\section{The equations of motion and the bulk Josephson effect}
Lagrange equations corresponding to (\ref{L}) are
\begin{eqnarray}
\label{LE}
i\frac{\partial \Psi_p}{\partial t}&=&
-\frac{\nabla^2_\alpha}{2m_p^\alpha}\Psi_p+2b\Psi_p
\left(\left|\Psi_p\right|^2-N_0/4\right)
+2e\varphi\Psi_p+|\gamma|\Psi_p + \gamma\Psi_{\bar p},\\
\Delta\varphi&=&-8\pi e\left(|\Psi_1|^2+|\Psi_2|^2-N_0/2\right),\nonumber
\end{eqnarray}
where ${\bar p}=2$ if p=1, and ${\bar p}=1$ if p=2.
To be specific let us consider the case $\gamma<0$ which corresponds to the 
d-wave ground state. Stiffness of the free energy (\ref{F}) with respect to  
variation of $|\Psi_p|$ is much larger than that with respect to the phases 
variation. Therefore we can represent the order parameters as 
$\Psi_p=\sqrt{N_0/4}(1 + \delta |\Psi_p|)\exp(i\phi_p)$, where 
$\delta|\Psi_p| \ll 1$, but the phase $\phi_p$ is arbitrary.
In the linear approximation in  $\delta|\Psi_p|$ the eqs.(\ref{LE}) can be 
rewritten as
\begin{eqnarray}
\label{LE1}
\frac{\partial \delta|\Psi_p|}{\partial t}&=&
-\frac{1}{2m_p^\alpha}\left(2\nabla_\alpha\phi_p\nabla_\alpha\delta|\Psi_p|+
\nabla^2_\alpha\phi_p\right)+|\gamma|\sin(\phi_p-\phi_{\bar p}),\nonumber \\
-\frac{\partial\varphi_p}{\partial t}&=&
-\frac{1}{2m^\alpha_p}\left(\nabla^2_\alpha\delta|\Psi_p|-
(\nabla_\alpha\phi_p)^2\right)+bN_0\delta|\Psi_p|+2e\varphi-
|\gamma|\left(\cos(\phi_p-\phi_{\bar p})-1\right),\\
\Delta \varphi&=&-4\pi e N_0(\delta|\Psi_1|+\delta|\Psi_2|).\nonumber
\end{eqnarray}
We will see below that $\delta|\Psi_1|+\delta|\Psi_2|=0$, i.e.
there is no  charge density. Therefore the scalar potential is 
due to the external field only: 
$\varphi=-{\bf E}\cdot {\bf r} \cdot \sin \omega t
=-(E_x x+ E_y y)\sin\omega t$, where ${\bf E}$ and $\omega$ are the amplitude 
and the frequency of the microwave field.
Let us introduce the phase $\phi_p^0$,
\begin{equation}
\label{phi0}
\phi_p^0=-\frac{2e}{\omega}{\bf E}\cdot {\bf r} \cdot
\cos\omega t-\frac{e^2}{2\omega^3}\left(\frac{E^2_{x}}{m_p^x}+\frac{
E^2_{y}}{m_p^y}\right)(2\omega t+\sin2\omega t),
\end{equation}
which satisfies the equation
\begin{equation}
\label{ss}
-\frac{\partial \phi^0_p}{\partial t}=\frac{1}{2m^\alpha_p}
\left(\nabla_\alpha \phi_p^0\right)^2-2e {\bf E} \cdot {\bf r} \cdot 
\sin(\omega t).
\end{equation}
After substitution $\phi_p=\phi_p^0+\delta\phi_p$ into eqs. (\ref{LE1}) 
we obtain equations for $\delta\phi_p$ and $\delta|\Psi_p|$. 
We are looking for $\delta\phi_p$ and $\delta|\Psi_p|$ independent of
coordinates and therefore the terms with spatial derivatives can be
omitted resulting in
\begin{eqnarray}
\label{22}
&&\frac{d}{dt}\left({\delta|\Psi_1|}+{\delta|\Psi_2|}\right)=0,\nonumber \\
&&\frac{d}{dt}\left({\delta|\Psi_2|}-{\delta|\Psi_1|}\right)=
2|\gamma|\sin\phi\\
&&\frac{d}{dt}({\delta\phi_2}+{\delta\phi_1})=-4|\gamma|\sin^2(\phi/2),
\nonumber\\
&&\frac{d}{dt}\left({\delta\phi_2}-{\delta\phi_1}\right)=
bN_0(\delta|\Psi_1|-\delta|\Psi_2|),\nonumber
\end{eqnarray}
where $\phi=\phi_2-\phi_1=\phi_2^0-\phi_1^0+\delta\phi_2-\delta\phi_1$.
First of eqs.(\ref{22}) proves that there is no induced charge density:
$\delta|\Psi_1|+\delta|\Psi_2|=0$.
According to eq. (\ref{phi0})
\begin{equation}
\label{AE}
\phi_2^0-\phi_1^0=-A(2\omega t+\sin 2\omega t),\hspace{0.2cm}
A=\frac{e^2}{2\omega^3}\left(E_y^2-E^2_x\right)\frac{M-m}{Mm}.
\end{equation}
From (\ref{22}) and (\ref{AE}) we find the equation for the relative phase $\phi$
\begin{equation}
\label{pend}
{\ddot \phi}+\omega_0^2 \sin \phi=A(2\omega)^2\sin 2\omega t,
\end{equation}
where $\omega_0=2\sqrt{|\gamma|a}$ is the threshold frequency of the 
phase excitation in the d-g-wave cuprates \cite{phase}. 
Note that the driving force in this equation reminds of that found in the
paper \cite{Rog}.

Equation (\ref{pend}) does not contain absorption. 
In the present work we follow the simplest phenomenological way of just 
adding a ``friction term'' $\Gamma {\dot \phi}$ to the left hand side of eq. 
(\ref{pend}) rewriting it as
\begin{equation}
\label{pend2}
{\ddot \phi}+\Gamma {\dot \phi}+\omega_0^2 \sin \phi=
A(2\omega)^2\sin 2\omega t.
\end{equation}
>From this equation one immediately derives  that absorption is proportional to 
\begin{equation}
\label{R}
R \propto \langle \Gamma {\dot \phi}^2\rangle \propto A\omega^3
\int_0^{2\pi/\omega} 
{\dot \phi} \sin(2\omega t) dt.
\end{equation}
The form of eq.(\ref{pend2}) is identical to the one describing the
resistively shunted Josephson junction which has been used 
in the work \cite{Zhai} for phenomenological description of
experimental data, see also Ref.\cite{Xie}.
We have derived this equation from the microscopic picture based on the
small Fermi surface. The derivation gives us all coefficients except the 
``friction'' which has been introduced phenomenologically.
This approach allows to find the dependence of the absorption upon
direction, amplitude and frequency of the external microwave field.
 
Concluding this section let us calculate the electric current.
\begin{equation}
\label{j}
j_{\alpha}={{2e}\over{m_1^{\alpha}}}|\Psi_1|^2\nabla_{\alpha}\phi_1+
{{2e}\over{m_2^{\alpha}}}|\Psi_2|^2\nabla_{\alpha}\phi_2.
\end{equation}
In the phase $\phi_p=\phi_p^0+\delta\phi_p$ only $\phi_p^0$ is
dependent on coordinates. Using eq. (\ref{phi0}) one finds
\begin{equation}
\label{j2}
{\bf j}=-{{(2e)^2}\over{\omega}}\left({1\over m}+{1\over M}\right)
\cdot {\bf E} \cdot \cos \omega t.
\end{equation}
This is the usual time dependent Meissner current.

\section{Numerical estimations and comparison with experiment}

Dimensionless form of eq. (\ref{pend}) is
\begin{equation}
\label{pend1}
\Lambda{\ddot \phi}+{\dot \phi}+\sin\phi=a\sin\Omega \tau,
\end{equation}
where $\tau=t\omega_0^2/\Gamma$, $\Lambda=\omega_0^2/\Gamma^2$,
$\Omega=2\omega \Gamma /\omega_0^2$, $a=4A\omega^2/\omega_0^2$, 
and the ``dot'' denotes the derivative over dimensionless time $\tau$.
Equation (\ref{pend1}) with $\Lambda=0$ (``overdamped pendulum'')
has been used in Ref.\cite{Zhai} to describe the experimental
data on coherent Josephson response of $YBaCu_2O_{7-\delta}$ single 
crystals at microwave frequency $\omega = 2\pi 10^{10}$.
The data has been fitted with $\Omega \approx 0.1$, and we take this value
as an experimental result.
Unfortunately our microscopic calculation does not give the value of $\Gamma$.
A reasonable fit of the experimental data \cite{Zhai} can be obtained with
$\Gamma \ge \omega_0$. To be specific we set $\Gamma= \omega_0$, then
$\omega_0=2\omega/\Omega=2\pi\cdot 200GHz$,
which is close to the estimation $\omega_0\sim2\pi\cdot 300GHz$ 
obtained in Ref. \cite{phase} from the microscopic picture.
Results of a numerical calculation of the normalized absorption
$R=\left<\dot{\phi}\sin\Omega\tau\right>\left/a \right.$ 
are shown in Fig.2. 
All features of the impedance data \cite{Zhai} are reproduced
and the first step  occurs at $|a|\approx 1.1$. (The critical value
of $|a|\approx 1$ is evident from eq. (\ref{pend1}) without any
computations.) This allows one to find the critical field $E_c$ - the field 
at which the first step occurs. Using eq. (\ref{AE}) we find
\begin{equation}
\label{thr}
\left|\frac{2e^2}{\omega \omega_0^2}\cdot \frac{M-m}{m M}
\cdot E_{c}^2\cos(2\beta)\right|=|a|\approx 1.
\end{equation}
Here $\beta$ is the angle between the field  and the axis x or y.
Taking $\cos(2\beta)=1$, reduced mass $mM/(M-m)$ equal to the electron mass,
and the frequencies $\omega=2\pi\cdot 10GHz$, $\omega_0=2\pi\cdot 200GHz$
we find the value of the critical field: $E_c \sim 450mGs$.
The experimental critical field is $H_c = 200mGs$ \cite{Zhai}.
This is in good agreement if we assume that at the surface of the sample 
the magnetic microwave field is of the order of the electric one.
Equation (\ref{thr}) determines the dependence of the critical field on
the microwave frequency and the angle between the field and the  x
axis
\begin{equation}
\label{EC} 
E_c \propto \sqrt{{{\omega}\over{|\cos(2\beta)|}}}.
\end{equation}
Note that $|\cos(2\beta)|=|\sin(2\theta)|$ where $\theta$ is the
angle between the electric microwave field and the crystal axes a or b.
\section{Conclusions}

We have considered the scenario with the small Fermi surface 
consisting of hole pockets.  The small Fermi surface together
with the mechanism of the magnetic pairing result in the possibility
of having two  superconducting condensates. Energy splitting
between these condensates is small and their linear combinations correspond 
to the d- and g-wave pairings. The ground state symmetry depends
on the interplay between the magnetic pairing and the Coulomb repulsion.
Microwave electric field applied in the ab-plane induces Josephson 
transitions between the condensates. This results in a nonlinear
absorption when the entire crystal responds like a single Josephson
junction. This could be an explanation of the effect observed recently
in $YBaCu_2O_{7-\delta}$ single crystals \cite{Zhai}.
The theoretical estimation for the critical microwave field 
obtained in the present work $H_c \sim E_c \sim 450mGs$
is close to the experimental value.
We predict  the dependence of the critical field on the microwave frequency 
and orientation of the microwave electric field with respect to the crystal
axes of the cuprate. This dependence is given by eq.(\ref{EC}).

\section{Acknowledgments}
We are grateful to S. Sridhar for communicating the result of work \cite{Zhai} 
prior to publication.
We acknowledge support from the Australian Research  Council.

\newpage

FIGURE CAPTIONS

\vspace{0.5cm}

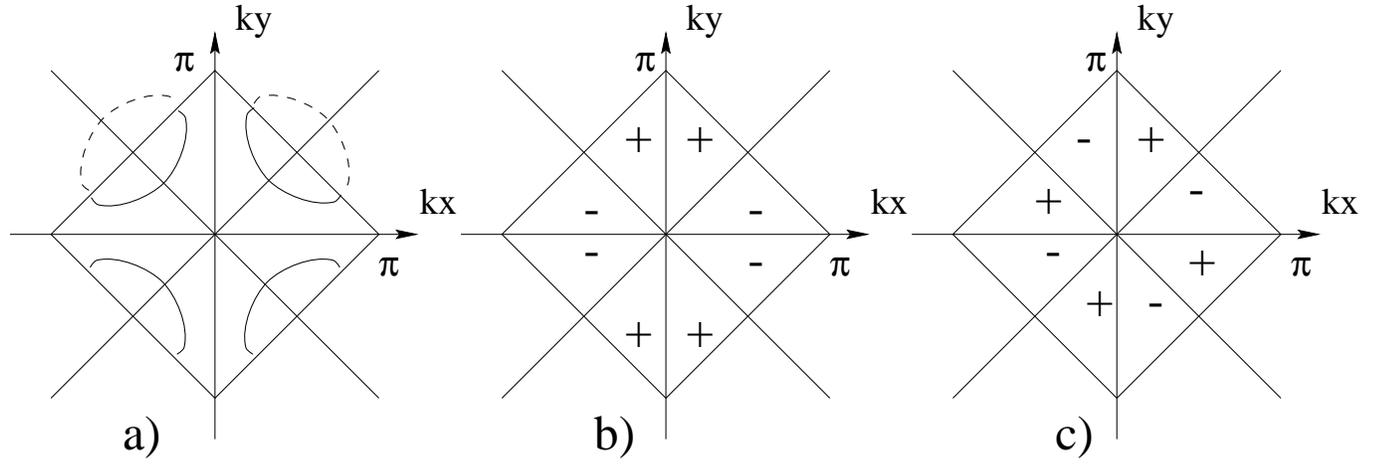
Fig.1.{\bf a}. Fermi surface in magnetic Brillouin zone which is equivalent 
to the two-pocket Fermi surface (dashed line). {\bf b}. Symmetry of the d-wave 
pairing in momentum space. {\bf c}. Symmetry of the g-wave pairing in momentum
 space.

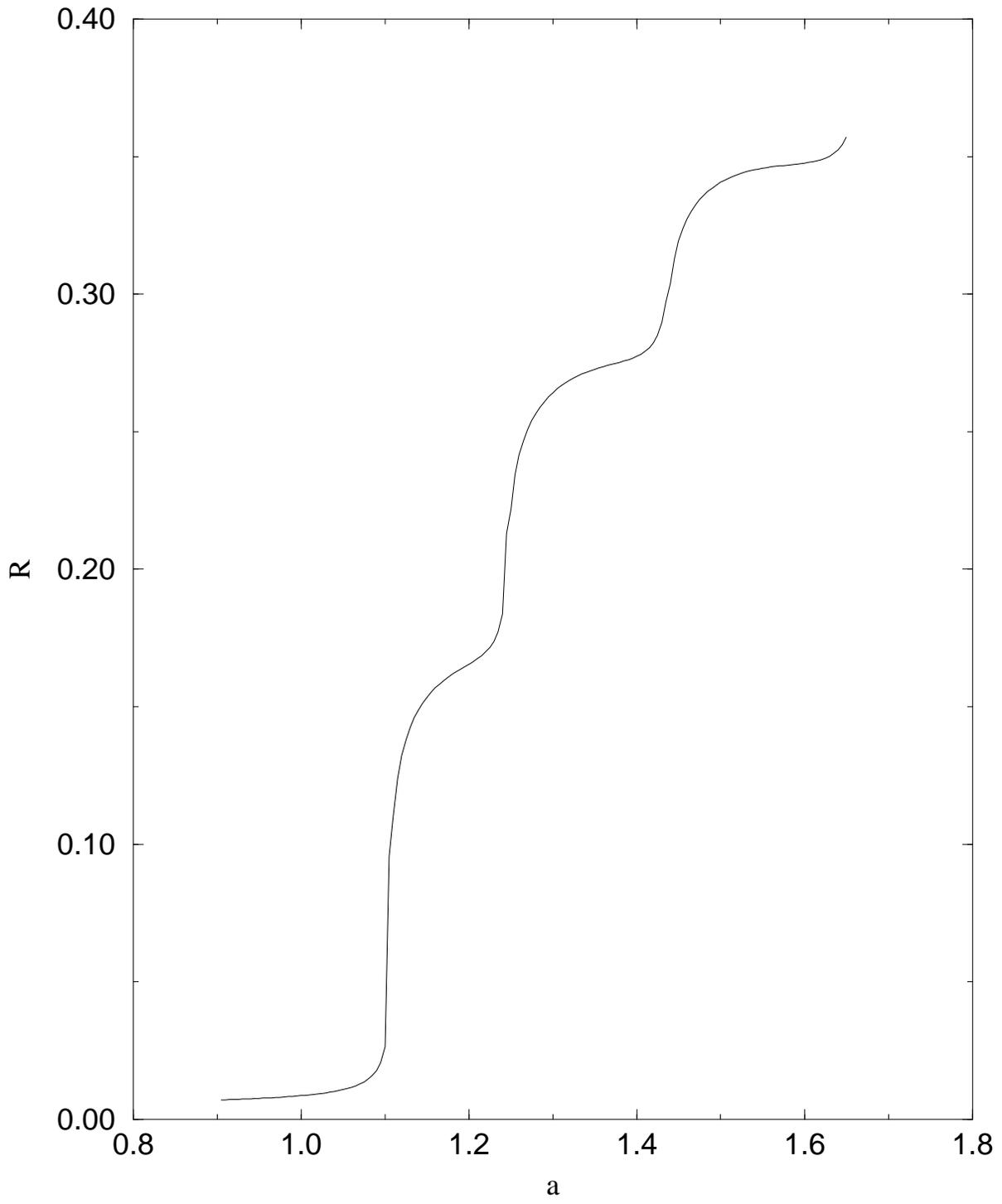
Fig.2. Normalized absorption $R$ versus dimensionless driving force $a$. 

\end{document}